\documentclass[twocolumn,showpacs,preprintnumbers,amsmath,amssymb]{revtex4}
\usepackage{amsmath}

\usepackage{graphicx}
\usepackage{dcolumn}
\usepackage{bm}

\begin{document}

\def\vsigma{{\hbox{\boldmath $\sigma$}}}
\def\vlambda{{\hbox{\boldmath $\lambda$}}}

\title{Characteristic quantities of pion-emitting sources extracted by
model-independent analysis in relativistic heavy ion collisions}

\author{Wei-Ning Zhang$^{1,2}$\footnote{wnzhang@dlut.edu.cn}}
\author{Zhi-Tao Yang$^2$}
\author{Yan-Yu Ren$^2$}

\affiliation{$^1$School of Physics and Optoelectronic Technology,
Dalian University of Technology, Dalian, Liaoning 116024, P. R. China\\
$^2$Department of Physics, Harbin Institute of Technology, Harbin,
Heilongjiang 150006, P. R. China\\
}

\begin{abstract}

We examine the characteristic quantities of pion-emitting sources
extracted by model-independent imaging analysis in relativistic
heavy ion collisions.  The moments of the spatial separation of pion
pair emission can provide the characteristic information about the
source geometry and coherence.  They are better for describing the
non-Gaussian sources with granular and core-halo structures.  An
improved granular source model of quark-gluon plasma droplets can
reproduce the main characteristics of the two-pion correlation
functions and source functions in the experiment of $\sqrt{s_{\rm
NN}}=200$ GeV Au+Au collisions.  The transverse-momentum dependence
of the normalized first-order moments of the separation for the
granular source is consistent with that of the usual interferometry
results of source radii, after taking into account the Lorentz
contraction in the direction of transverse momentum of pion pair.
\end{abstract}


\maketitle

\section{Introduction}

Hanbury-Brown-Twiss (HBT) interferometry has been extensively used
to extract the space-time and coherence information about the
particle-emitting sources produced in relativistic heavy ion
collisions \cite{CYW94,UAW99,RMW00,MAL05}.  In conventional two-pion
HBT analysis one needs fitting the measured correlation functions
with parameterized formulae, e.g., a Gaussian form, to obtain
quantitatively the source radii and chaotic parameter.  So these
quantitative HBT results are model depended.  Recently, many studies
indicate that the particle-emitting sources produced in relativistic
heavy ion collisions are far from Gaussian distributed
\cite{DAB01,PCH05,PHE07,PCH07,PCH08,PHE08,RAL08,ZTY09,ZWL02,ZWL04,TCS04,WNZ06,YYR08}.
For the non-Gaussian sources the conventional HBT method of the
Gaussian fit is inappropriate \cite{SNI98,UHE99,DHA00,TCS04,EFR06}.
Therefore, it is important to investigate the source characteristic
quantities by model-independent analysis.

The imaging technique introduced by Brown and Danielewicz
\cite{DAB97,DAB98,DAB01} is a model-independent way to obtain the
two-pion source function $S(r)$, the probability for emitting a pion
pair with spatial separation $r$ in the pair center-of-mass system
(PCMS), from the measured two-pion correlations.  This technique has
been developed and used in analyzing one- and multi-dimension source
geometry in relativistic heavy ion collisions
\cite{DAB00,DAB01,SYP01,GVE02,PCH03,PDA04,PDA05,DAB05,PCH05,PDA07,PHE07,PCH07,PCH08,PHE08,RAL08,ZTY09}.
Recent imaging analyses for the pion-emitting sources produced in
relativistic heavy ion collisions suggest that the sources may have
a core-halo \cite{PCH05,PHE07} or granular structure \cite{ZTY09}.
In this paper we will examine the characteristic quantities
extracted by the imaging analysis for the non-Gaussian sources with
granular and core-halo structures.  We find that the zero-order
moment of $r$, corresponding to the intercept of the two-pion
correlation function at zero relative momentum, is usually larger
than the chaotic parameter obtained by the HBT Gaussian fit. The
first-order moment and standard deviation of $r$, which are
normalized to the Gaussian radius for a Gaussian source, can be used
to characterize the source size and the deviation of the source
distribution from Gaussian form.

In Ref. \cite{WNZ06}, a granular source model was used to explain
the HBT puzzle, $R_{\rm out}/R_{\rm side}\sim 1$, in the
Relativistic Heavy Ion Collider (RHIC) experiments
\cite{STA01a,PHE02a,PHE04a,STA05a}.  Here $R_{\rm out}$ and $R_{\rm
side}$ are the HBT radii in the directions parallel and
perpendicular to the transverse momentum of pion pair
\cite{GBE88,SPR90}.  However, in the previous granular source models
\cite{WNZ04,WNZ06}, all the droplets are assumed with the same
initial radius and pions are emitted at a fixed freeze-out
temperature $T_f$.  In this paper we will improve the granular
source model by introducing random initial radii of the droplets
according to a Gaussian distribution and letting the pions emit in a
wide $T_f$ region for including both directly produced and decayed
pions.  We will investigate the characteristic quantities extracted
by the three-dimension imaging analysis \cite{PDA05,PDA07} for the
improved granular source.  We find that the improved granular source
mode can reproduce the main characteristics of the experimental
two-pion correlation functions and source functions in $\sqrt{s_{\rm
NN}}=200$ GeV Au+Au collisions \cite{PHE04a,STA05a,PHE08}.  The
transverse-momentum dependence of the normalized first-order moments
agrees with that of the HBT radii obtained by usual Gaussian fit,
after taking into account the Lorentz contraction in the direction
of transverse momentum of pion pair.

The paper is organized as follows.  In section II, we discuss
briefly the limitation of the HBT Gaussian fit for the non-Gaussian
sources with granular and core-halo structures.  The
model-independent characteristic quantities extracted by imaging
analysis are examined for the non-Gaussian sources and compared with
the HBT results of the Gaussian fit.  In section III we introduce an
improved granular source model of QGP droplets to simulate the
two-pion HBT correlations in relativistic heavy ion collisions.  In
section IV we investigate the three-dimension source functions for
the improved granular source.  The transverse-momentum dependence of
the characteristic quantities of the source functions are
investigated in different directions.  Finally, a summary and
conclusion is given in section V.

\section{Gaussian fit and imaging analysis in HBT interferometry}

In this section we discuss the quantitative results extracted by the
usual Gaussian fit and a model-independent imaging analysis in HBT
interferometry.  For a direct comparison with analytical results, we
consider in this section only the static sources with spherical
symmetry.

\subsection{Gaussian fit}

In conventional HBT analysis, the measured two-pion correlation
functions are fitted with a Gaussian formula to obtain quantitative
results.  Using the relative-momentum $q$ of the pair as a variable,
the Gaussian fitting formula is
\begin{eqnarray}
\label{CqG} C(q)=1+{\cal R}(q)=1+\lambda_{\rm hbt}\exp(-q^2R_{\rm
hbt}^2)\,,
\end{eqnarray}
where $R_{\rm hbt}$ and $\lambda_{\rm hbt}$ are called the HBT
radius and chaotic parameter of the pion-emitting source.

Let us examine the fitting procedure.  The chi-square function is
\cite{EFR06}
\begin{eqnarray}
\label{chi2} \chi^2\equiv\sum_i\left[\frac{\ln{\cal
R}_i-\ln\lambda_{\rm hbt}+q_i^2R_{\rm hbt}^2}{\sigma_i'}\right]^2\,,
\end{eqnarray}
where ${\cal R}_i$ is the $i$th measured correlator at $q_i$,
$\sigma_i'={\sigma_i}/{{\cal R}_i}$, and $\sigma_i$ is the error of
${\cal R}_i$.  Minimizing the $\chi^2$ with respect to the fitting
parameters $R_{\rm hbt}^2$ and $\ln\lambda_{\rm hbt}$,
\begin{eqnarray}
\frac{\partial\chi^2}{\partial(R_{\rm hbt}^2)}=0\,,\ \ \ \ \
\frac{\partial\chi^2}{\partial(\ln\lambda_{\rm hbt})}=0\,,
\end{eqnarray}
one obtains
\begin{eqnarray}
\label{RGF}
R_{\rm hbt}^2= \frac{\sum\limits_{i,j}
a_ia_j(b_i-b_j)q_i^2}{\sum\limits_{i,j}^{} a_i a_j
(q_i^2-q_j^2)q_i^2}\,,
\end{eqnarray}
\begin{eqnarray}
\label{lGF}
\ln\lambda_{\rm hbt}=-\frac{\sum\limits_{i,j} a_i a_j
\,b_j(q_i^2-q_j^2)q_i^2}{\sum\limits_{i,j}^{} a_i a_j
(q_i^2-q_j^2)q_i^2}\,,
\end{eqnarray}
where the quantities
\begin{eqnarray}
a_i={\cal R}_i^2/\sigma_i^2\,,\ \ \ \ \ b_i=\ln{\cal R}_i\,.
\end{eqnarray}
Introduce
\begin{equation}
A=\sum_j a_j,~~~~B=\sum_j a_j\, b_j,
\end{equation}
\begin{equation}
C=\sum_j a_j\,b_j\,q_j,~~~~D=\bigg|\sum_{i,j} a_i a_j
(q_i^2-q_j^2)q_i^2 \bigg|,
\end{equation}
equations (\ref{RGF}) and (\ref{lGF}) can be expressed as
\begin{equation}
R_{\rm hbt}^2 = \frac{1}{D} \bigg| \sum_i (A b_i -B)\,a_i \,q_i^2
\bigg|,
\end{equation}
\begin{equation}
\ln\lambda_{\rm hbt} = -\frac{1}{D} \bigg| \sum_i (B q_i^2 -C)\,a_i
\,q_i^2 \bigg|.
\end{equation}
It can be seen that the term for the contribution of the $i$th
measured correlator has a weight factor of ${\cal R}_i^2
q_i^2/\sigma_i^2$.  At a smaller $q_i$, ${\cal R}_i \sim 1$ and
$\sigma_i$ is usually larger than that for a larger $q_i$.  So the
contributions from the data at the bins near zero $q$ are very
small.

Recent imaging analyses in relativistic heavy ion collisions exhibit
a two-tiered structure or a long-tail in the two-pion source
functions \cite{PHE07,PCH07,PCH08,PHE08,RAL08,ZTY09}.  These
non-Gaussian source functions indicate that the pion-emitting
sources may have a granular \cite{ZTY09} or core-halo
\cite{PCH05,PHE07} structure. Let us examine the Gaussian-fit to the
HBT correlation functions of these kinds of non-Gaussian sources.

For the granular source, the pions are assumed to be emitted from
separated droplets in the source
\cite{SPR92,WNZ95,WNZ04,WNZ06,ZTY09}.  Assuming that the
distributions of the emitting-points in each droplet and the droplet
centers in the source are Gaussian forms with standard deviations
$a$ and $R_{\rm gr}$ respectively, the two-pion correlation function
for a static granular source is given by \cite{SPR92,WNZ95}
\begin{eqnarray}
\label{CqGR}
C_{\rm gr}(q)=1+\frac{1}{N}e^{-q^2a^2}+
\Big(1-\frac{1}{N}\Big) e^{-q^2(a^2+R_{\rm gr}^2)},
\end{eqnarray}
where $N$ is the number of droplet in the source.  In Eq.
(\ref{CqGR}) the second term is the contribution corresponding to
the two pions from the same droplet and the third term is for the
pions from different droplets.

For the core-halo source \cite{TCS96,SNI98}, the pions are emitted
from a central core and a halo of long-lived resonance decays.
Assuming the emitting-points in the core and halo are all the
Gaussian distribution with standard deviations $R_{\rm c}$ and
$R_{\rm h}$, the two-pion correlation function of the static
core-halo source is given by \cite{SNI98}:
\begin{eqnarray}
\label{CqCH}
C_{\rm ch}(q)&=&1+f_{\rm c}^2\,e^{-q^2R_{\rm
c}^2}+(1-f_{\rm c}^2)e^{-q^2R_{\rm h}^2}\nonumber\\
&&+ 2f_{\rm c}(1-f_{\rm c})\,e^{-q^2(R_{\rm c}^2+R_{\rm h}^2)/2},
\end{eqnarray}
where $f_{\rm c}$ is the fraction of core emission.  The second,
third, and fourth terms in Eq. (\ref{CqCH}) are the contributions
for both the two pions from the core, the halo, and one from the
core and another from the halo, respectively.

\begin{figure} [h]
\includegraphics[angle=0,scale=0.4]{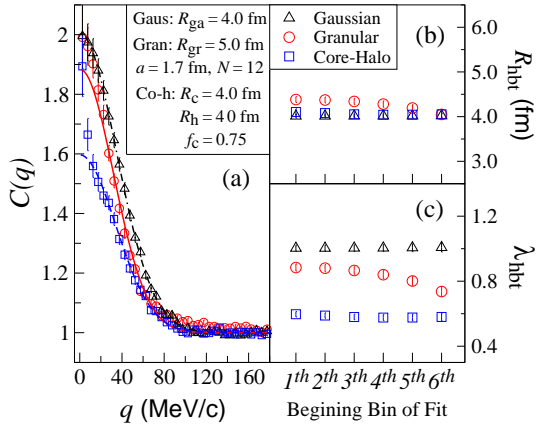}
\caption{(Color online) (a) Two-pion correlation functions for
completely chaotic Gaussian, granular, and core-halo sources.  (b)
and (c) The fitted results of $R_{\rm hbt}$ and $\lambda_{\rm hbt}$
of the Gaussian fit with the data sets beginning from the $i$th
measured correlator. } \label{Figgafit}
\end{figure}

In Fig. \ref{Figgafit}(a) we show the simulated two-pion
correlations for completely chaotic Gaussian, granular, and
core-halo sources, where the lines are the fitted curves of the
Gaussian fit, Eq. (\ref{CqG}).  It can be seen that the fit is very
good for the Gaussian source.  However, for the granular and
core-halo sources the fits are inappropriate, which miss the data of
the lower $q$-bins.  In Fig. \ref{Figgafit}(b) and (c), we exhibit
the fitted results of $R_{\rm hbt}$ and $\lambda_{\rm hbt}$ for the
fits which use the data sets beginning from the $i$th measured
correlator.  One can see that the fitted results of $R_{\rm hbt}$
are insensitive to the first several data of $q$-bins.  For the
core-halo source, the $R_{\rm hbt}$ results are almost the same as
that of the Gaussian source.  The fitted HBT radii cannot reflect
the large spatial extent of the halo for the long-lived resonance
decays.  Also, because of missing the data of the first one or two
$q$-bins, the results of $\lambda_{\rm hbt}$ for the core-halo
source are about $f_{\rm c}^2$, which is the main reason that a
large halo-structure was introduced to explain the low chaotic
parameters measured in experiments \cite{TCS96,SNI98}.  For the
granular source, the two-tiered structure of the correlations
\cite{SPR92,WNZ95} and the failure for the first several data in the
Gaussian fit lead to $R_{\rm hbt} < R_{\rm gr}$ and $\lambda_{\rm
hbt} <1 $.  So, the usual model-dependent HBT fitted results are
hardly to reflect the full geometry and coherence of these
non-Gaussian sources.

\subsection{Imaging analysis}

The imaging technique introduced by Brown and Danielewicz
\cite{DAB97,DAB98,DAB01} allows one to obtain the two-pion source
function $S(r)$ in PCMS, from the measured two-pion correlation
function $C(Q)~(Q=\sqrt{{\bf q}^2 -q_0^2})$. After knowing the
source function one can calculate the moments of $r$
model-independently. Define the $n$th-order moment of $r$ as:
\begin{eqnarray}
\label{rn}
\langle r^n\rangle=\frac{4\pi\int_0^\infty
dr\,r^n\,S(r)\,r^2}{4\pi\int_0^\infty
dr\,S(r)\,r^2}\,,~~~n=1,2,\cdots,
\end{eqnarray}
where the denominator is a quantity of normalization as the
zero-order moment of $r$.  Denote the zero-order moment of $r$ by
$\widetilde{\lambda}$,
\begin{eqnarray}
\label{wl} \widetilde{\lambda} \equiv 4\pi\int_0^\infty drS(r)r^2\,.
\end{eqnarray}
Theoretically, the value of $\widetilde{\lambda}$ is equal to ${\cal
R}(Q=0)$ \cite{DAB01}.  It is unit for a completely chaotic source
and between zero and unit for a partially coherent source.

For a completely chaotic Gaussian source, the source function is
\cite{DAB98,DAB01,SYP01}
\begin{eqnarray}
\label{SGA} S_{\rm ga}(r)=\frac{1}{(\sqrt{4\pi} R_{\rm ga})^3}
\exp\left(-\frac{r ^2}{4R_{\rm ga}^2}\right)\,,
\end{eqnarray}
where $R_{\rm ga}$ is called as the Gaussian radius of the source.
The first- and second-order moments of $r$ for the Gaussian source
are:
\begin{eqnarray}
\label{Gam}
\langle r\rangle_{\rm ga}=\frac{4}{\sqrt{\pi}}R_{\rm
ga}\,,~~~~\langle r^2\rangle_{\rm ga}=6R_{\rm ga}^2\,.
\end{eqnarray}
The deviation of the Gaussian source function is
\begin{eqnarray}
\sigma=\sqrt{\langle r^2\rangle-\langle r\rangle^2}=\sqrt{6-16/\pi}
R_{\rm ga}\,.
\end{eqnarray}

The source functions for completely chaotic granular and core-halo
sources are given by \cite{ZTY09,ZTY09c}
\begin{eqnarray}
\label{SGR} S_{\rm gr}(r)&=&
\frac{1}{N}\frac{1}{(\sqrt{4\pi}a)^3}\exp\left(-\frac{r^2}{4a^2}\right)
+(1-\frac{1}{N})\notag\\
&\times& \!\!\frac{1}{[4\pi(a^2\!+\!R_{\rm
gr}^2)]^{3/2}}\exp\!\!\left[\!-\frac{r^2}{4(a^2\!+\!R_{\rm
gr}^2)}\right],
\end{eqnarray}
\begin{eqnarray}
\label{SCH} S_{\rm ch}(r)&=&
\frac{f_c^2}{(\sqrt{4\pi}
R_c)^3}\exp\left(-\frac{r^2}{4R_c^2}\right)\notag\\
&+&\!\!\frac{(1-f_c)^2}{(\sqrt{4\pi}
R_h)^3}\exp\left(-\frac{r^2}{4R_h^2}\right)\notag\\
&+&\!\!\frac{2f_c(1-f_c)}{[2\pi(R_c^2\!+\!R_h^2)]^{3/2}}\exp\!\!
\left[\!- \frac{r^2}{2(R_c^2\!+\!R_h^2)}\right].
\end{eqnarray}
In Eq. (\ref{SGR}) the first term is the contribution corresponding
to the two pions from the same droplet and the second term is for
the two pions from different droplets.  In Eq. (\ref{SCH}) the first
and second terms are the contributions for the two particles both from
the core and from the halo, and the third term corresponds to that one
particle from the core and another from the halo.

We introduce a normalized first-order moment and a normalized
deviation as
\begin{equation}
\label{wR}
\widetilde{R} \equiv \frac{\sqrt{\pi}}{4}\langle r\rangle
\end{equation}
and
\begin{equation}
\label{wsigma} \widetilde{\sigma} \equiv \sigma /\sqrt{6-16/\pi}\,.
\end{equation}
Both the quantities are normalized to the Gaussian radius $R_{\rm
ga}$ for a Gaussian source.  The normalized moment $\widetilde{R}$
describes the source size.  It is agrees with the HBT radius for a
Gaussian source and we shall see that it is greater than the HBT
radii for the granular or core-halo source.  The normalized
deviation $\widetilde{\sigma}$ is equal to $\widetilde{R}$ for a
Gaussian source and larger than $\widetilde{R}$ for the granular or
core-halo source.  One can examine the deviation of source
distribution from Gaussian distribution by comparing the values of
$\widetilde{R}$ and $\widetilde{\sigma}$.

\vspace*{3mm}
\begin{figure} [h]
\includegraphics[scale=0.55]{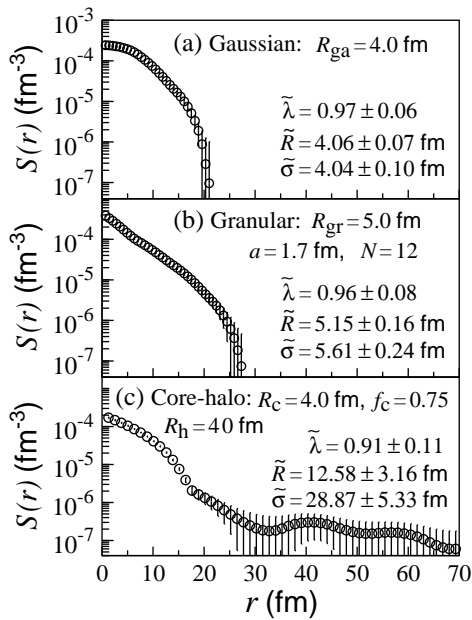}
\vspace*{1mm}\caption{The two-pion source functions for the
Gaussian, granular, and core-halo sources as the same in figure
\ref{Figgafit}.} \label{Figimggc}
\end{figure}

In Fig. \ref{Figimggc} we show the source functions obtained by
imaging analysis for the Gaussian, granular, and core-halo sources
as in Fig. \ref{Figgafit}.  The values of $\widetilde{\lambda}$,
$\widetilde{R}$, and $\widetilde{\sigma}$ are calculated with the
imaging source functions.  They are consistent with the
corresponding analytical results obtained by Eqs. (\ref{SGA}),
(\ref{SGR}) -- (\ref{wsigma}).  It can be seen that the source
function of the granular source is wider than that of the Gaussian
source.  Because of the long-lived resonance halo the source
function of the core-halo source has a very long tail.  By comparing
with the results of the Gaussian fits in Fig. \ref{Figgafit}, one
can see that the results of $\widetilde{\lambda}$ and
$\widetilde{R}$ for the Gaussian source are consistent with the
corresponding HBT fitted results of $\lambda_{\rm hbt}$ and $R_{\rm
hbt}$, respectively.  The value of $\widetilde{\sigma}$ for the
Gaussian source is equal to that of $\widetilde{R}$ as expected. For
the non-Gaussian sources, the results of $\widetilde{\lambda}$
calculated with the source functions are close to unit and give more
realistic chaotic degrees of the sources than that given by the
$\lambda_{\rm hbt}$ results.  The results of $\widetilde{R}$ for the
non-Gaussian sources are larger than the corresponding $R_{\rm hbt}$
results. The first-order moment $\widetilde{R}$ reflects more about
the whole sizes of these non-Gaussian sources than $R_{\rm hbt}$,
because the Gaussian fit is insensitive to the measured data in
smaller relative momentum region (corresponding to larger coordinate
region).  The results of $\widetilde{\sigma} > \widetilde{R}$ for
the non-Gaussian sources reflect the deviations of the source
distributions from Gaussian form.

The values of $\widetilde{R}$ and $\widetilde{\sigma}$ are related
to the space-structure of the source.  For the granular source the
quantity $\xi=[(R_{\rm gr}/a)^3/(N-2)]~(1<N<R_{\rm gr}^3/a^3)$
characterizes the granularity of the source sensitively
\cite{ZTY09}. In Table \ref{tabimgr} we list the values of $\xi$ as
well as the results of $\widetilde{R}$ and $\widetilde{\sigma}$
calculated with the imaging source functions for the granular
sources with different source parameters.  The values of
$\widetilde{R}$ and $\widetilde{\sigma}$ change with the droplet
radius $a$ and number $N$ for the sources with the same radius
$R_{\rm gr}$.  The difference between the values of $\widetilde{R}$
and $\widetilde{\sigma}$ increases with $\xi$.  The
model-independent observables $\widetilde{R}$ and
$\widetilde{\sigma}$ provide the spatial information of the granular
sources.

\begin{table}
\caption{\label{tabimgr}The characteristic quantities of the
granular sources ($R_{\rm gr}=5.5$ fm).}
\begin{ruledtabular}
\begin{tabular}{cccc}
source parameters&$\xi$&$\widetilde{R}\,({\rm fm})$&$\widetilde{\sigma}\,
({\rm fm})$\\
\hline\\[-1ex]
$a=1.2\,$fm, $N=12$&9.63&5.32 $\pm$ 0.13&6.04 $\pm$ 0.16\\
$a=1.7\,$fm, $N=12$&3.39&5.47 $\pm$ 0.12&6.05 $\pm$ 0.15\\
$a=1.7\,$fm, $N=20$&1.88&5.61 $\pm$ 0.13&5.91 $\pm$ 0.17\\
\end{tabular}
\end{ruledtabular}
\end{table}

\section{An improved granular source model}

As it is seen in section II, the source functions of the granular
and core-halo sources are wider than that of the Gaussian source,
which agrees with recent imaging analyses of RHIC experiments
\cite{PCH05,PHE07,ZTY09}.  Because granular source model can
reproduce the experimental results of HBT radii \cite{WNZ06}, we
construct here an evolution pion-emitting source based on granular
source model and take into account the effect of resonance decay by
letting pions emit in a wide temperature region.

On the basis of Bjorken picture \cite{JDB83} the systems produced in
relativistic heavy ion collisions reach a local-equilibrium at
$\tau_0 \sim 1$ fm/c, then expand hydrodynamically.  The expanding
velocity of a fluid cell in a central rapidity region $|y|<y_m$ at
coordinate point $(r_{\perp}\equiv \rho,z,t)$ can be expressed as
\cite{GBA83}
\begin{equation}
\label{vcell}
v_{\perp}=v_{\perp}(z=0)\sqrt{1-v_z^2},~~~~v_z=z/t.
\end{equation}

We assume that the system fragments and forms a granular source of
many QGP droplets at a time $t_0 (>\tau_0=1~{\rm fm/c})$.  The
fragmentation may be due to the violent expansion of the system with
large fluctuation of initial matter distribution \cite{WNZ06} or the
rapidly increased bulk viscosity in the QGP near the phase
transition \cite{GTO08}.  Because of the surface tension of the QGP,
the droplet has a spherical geometry in its local frame. We assume
that the initial radii $r'_0$ of the droplets in droplet local frame
have a Gaussian distribution with standard deviation $a$, and the
initial droplet centers are distributed within a short cylinder
along the beam direction ($z$ direction) with the probabilities
\cite{WNZ06}
\begin{eqnarray}
\frac{dP_{\perp}}{2\pi\rho_0\,d\rho_0} \propto
\left[1-\exp\,(-\rho_0^2/\Delta{\cal
R}_{\perp}^2)\right]\theta({\cal R}_{\perp}-\rho_0)\,,
\end{eqnarray}
\begin{eqnarray}
\frac{dP_y}{dy_0}=\theta(y_m-|y_0|),~~~~z_0=t_0 \tanh y_0,
\end{eqnarray}
where $\rho_0$ and $z_0$ are the initial transverse and longitudinal
coordinates of the droplet, $y_0$ is the initial rapidity of the
droplet, and ${\cal R}_{\perp}$ and $\Delta{\cal R}_{\perp}$ are the
initial transverse radius and shell parameter of the granular
source.  In our calculations we take ${\cal R}_{\perp}=6.8$ fm,
$\Delta{\cal R}_{\perp}=3.3$ fm, and $a=2.3$ fm.  The initial
temperature of the droplets is taken bo be 200 MeV and the
quantities $y_m$ and $t_0$ are taken to be 1.5 and 5.0 fm/c.

The velocity of a droplet dependents on its initial
central-coordinate $(\rho_0, z_0)$.  Based on Eq. (\ref{vcell}) we
assume that the velocity of the droplet is given by
\begin{equation}
\label{vdrop} v_{d \perp}=a_T \Big(\frac{\rho_0}{{\cal
R}_{\perp}}\Big)^{b_T}\sqrt{1-v_{dz}^2}\,,~~~~v_{dz}=z_0/t_0,
\end{equation}
where $a_T$ and $b_T$ are the magnitude and exponential power
parameters determined by the particle transverse momentum
distribution.

The evolution of the system after $t_0$ is the superposition of all
the evolutions of the individual droplets, each of them is described
by relativistic hydrodynamics with the equation of state (EOS) of
the entropy density \cite{JPB87,ELA96,DHR96}.  The values of the EOS
parameters we used are the transition temperature $T_c=170$ MeV, the
transition temperature width $\Delta T_c=0.05 T_c$, the bag constant
$B=250\,{\rm MeV/fm^3}$, and the ratio of the degrees of freedom of
the QGP to the hadronic gas $d_Q/d_H=3.6$.

In order to include the pions emitted directly at hadronization and
decayed from resonances later, we let the pions freeze-out (emit out
from the source) within a wide temperature region with the
probability
\begin{eqnarray}
\label{Pt}
\frac{dP_f}{dT} &\propto& f_{\rm dir}
\exp[-(T-T_h)/\Delta T_{\rm dir}] \nonumber \\
&+&(1-f_{\rm dir}) \exp[-(T-T_h)/\Delta T_{\rm dec}]\,,
\end{eqnarray}
where $f_{\rm dir}$ is a fraction parameter for the direct emission,
$T_h$ represents the temperature of complete hadronization, $\Delta
T_{\rm dir}$ and $\Delta T_{\rm dec}$ describe the widths of
temperature for the direct and decayed pion emissions.

\begin{figure} [h]
\vspace*{-8mm}
\includegraphics[scale=0.75]{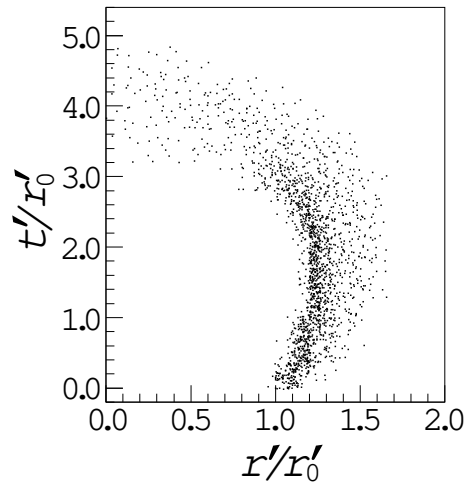}
\vspace*{-2mm} \caption{The space-time distribution of pion-emitting
points in the center-of-mass frame of droplet.} \label{Figdisstd}
\end{figure}

Figure \ref{Figdisstd} shows the space-time distribution of
pion-emitting points in the center-of-mass frame of droplet.
Considering that most pions are emitted directly from the
hadronization configuration and the decayed pions are produced
within a larger temperature region, we take the parameters in Eq.
(\ref{Pt}) to be $f_{\rm dir}=0.85$, $\Delta T_{\rm dir}=10$ MeV,
and $\Delta T_{\rm dec}=90$ MeV in our calculations.  The complete
hadronization temperature $T_h$ is taken to be $0.9T_c$.

\begin{figure} [h]
\vspace*{5mm}
\includegraphics[scale=0.5]{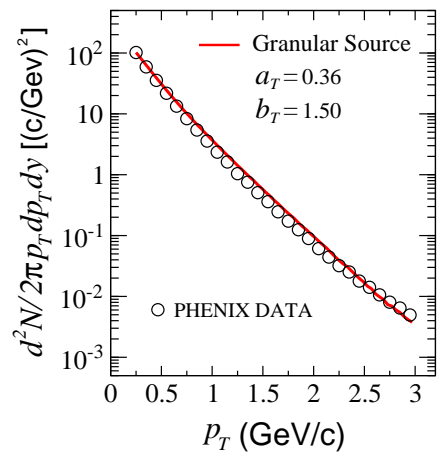}
\vspace*{2mm} \caption{(Color online) The pion transverse momentum
distribution of the granular source and the PHENIX data for
$\sqrt{s_{\rm NN}}=200$ GeV Au+Au collisions with minimum bias
\cite{PHE04}.} \label{Figsp}
\end{figure}

\begin{figure} [h]
\vspace*{1mm}
\includegraphics[scale=0.5]{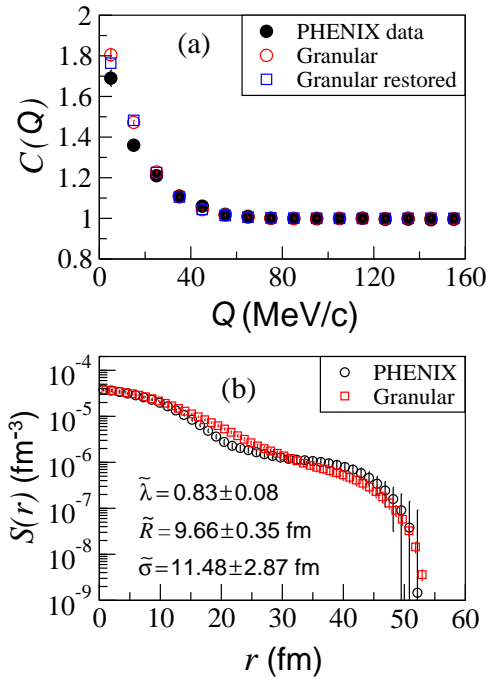}
\vspace*{5mm} \caption{(Color online) (a) The two-pion correlation
functions of the granular source and PHENIX experiment
\cite{PHE04aa}.  (b) The imaging results corresponding to the
correlation functions in (a).} \label{Figim1d}
\end{figure}

In Fig. \ref{Figsp}, we exhibit the pion transverse momentum
distribution of the granular source.  By comparing with the
experimental data \cite{PHE04}, we determine the velocity parameters
of the droplet in Eq. (\ref{vdrop}) as $a_T=0.36$ and $b_T=1.50$.
In Figure \ref{Figim1d} (a) we show the two-pion correlation
functions $C(Q)$ of the granular source (open circle) and PHENIX
experiment $\sqrt{s_{\rm NN}}=200$ GeV Au+Au collisions (solid
circle) \cite{PHE04aa}.  In Fig. \ref{Figim1d} (b) we show the
source functions $S(r)$ corresponding to the correlation functions
in Fig. \ref{Figim1d} (a).  It can be seen that the source function
of the granular source exhibits a long tail as the experimental
result. The symbols {\raisebox{-1.2mm} {\textsuperscript{$\Box$}}}
in Fig. \ref{Figim1d} (a) are for the restored correlation function
of the granular source. It is consistent with the original
correlation function of the granular source.  The difference between
the granular and experimental correlation functions at small $Q$ may
be arising from the final state interaction and the coherence of
pion emission which we have not considered in our granular source
model.

\section{Source HBT characteristic quantities}

In this section we examine the three-dimension imaging of the
improved granular source with the technique proposed by Danielewicz
and Pratt \cite{PDA05,PDA07}.  We will investigate the
characteristic quantities $\widetilde{\lambda}$, $\widetilde{R}$,
and $\widetilde{\sigma}$ in different directions as well as the
usual HBT radii $R_{\rm out}$, $R_{\rm side}$, and $R_{\rm long}$
for the granular source.

Using a Cartesian harmonic basis $\{{\cal
A}^l_{\alpha_{\!1},\cdots,\alpha_l}(\Omega)\}~(l=0,1,2,\cdots;\,\alpha_i=
x,y,\, {\rm or}\,z)$ \cite{PDA05,PDA07}, the three-dimension
correlation function ${\cal R}({\bf Q}\,)=C({\bf Q}\,)-1$ and source
function $S({\bf r})$ in PCMS can be expressed as
\cite{PDA05,PDA07,PHE08}
\begin{equation}
\label{RQ} {\cal R}({\bf Q}\,)=\sum_{l,\alpha_{\!1},\cdots,\alpha_l}
R^l_{\alpha_{\!1},\cdots,\alpha_l}(Q) {\cal
A}^l_{\alpha_{\!1},\cdots,\alpha_l}(\Omega_{\bf Q}),
\end{equation}
\begin{equation}
\label{S3r}
S({\bf r})=\sum_{l,\alpha_{\!1},\cdots,\alpha_l}
S^l_{\alpha_{\!1},\cdots,\alpha_l}(r) {\cal
A}^l_{\alpha_{\!1},\cdots,\alpha_l}(\Omega_{\bf r}),
\end{equation}
where ${\bf Q}$ and ${\bf r}$ are the relative momentum and
coordinate of the pion pair in PCMS, $Q$ and $\Omega_{\bf Q}$ are
the modulus and solid angle of ${\bf Q}$, and $r$ and $\Omega_{\bf
r}$ are the modulus and solid angle of ${\bf r}$. From the
Koonin-Pratt formalism \cite{SEK77,SPR90}, the relation of
$R^l_{\alpha_{\!1},\cdots,\alpha_l}(Q)$ and
$S^l_{\alpha_{\!1},\cdots,\alpha_l}(r)$ is given by
\cite{PDA05,PDA07,PHE08}
\begin{equation}
\label{RS} R^{\,l}_{\alpha_{\!1},\cdots,\alpha_l}(Q) =4\pi \int
dr\,r^2 {\cal K}_l(Q,r) S^l_{\alpha_{\!1},\cdots,\alpha_l}(r),
\end{equation}
where \cite{DAB05,PDA07}
\begin{equation}
{\cal K}_l(Q,r)=\left\{
\begin{array}{ll}
(-1)^{l/2} j_{\,l}(Qr) \, ,&\mbox{for even $l$} \, ,\\
0 \, ,&\mbox{for odd $l$} \, , \end{array} \right.
\end{equation}
if neglecting the final state interaction of the pion pair.

From Eq. (\ref{RQ}) one can get $
R^l_{\alpha_{\!1},\cdots,\alpha_l}(Q)$ from the measured
three-dimension correlation function ${\cal R}({\bf Q}\,)$
\cite{PDA05,PDA07,PHE08},
\begin{eqnarray}
\!\!\!\!R^{\,l}_{\alpha_{\!1},\cdots,\alpha_l}(Q)\!
=\!\frac{(2l+1)!!}{l!}\!\! \int\! \frac{d\Omega_{\bf Q}}{4\pi} {\cal
A}^l_{\alpha_{\!1},\cdots,\alpha_l}(\Omega_{\bf Q}) {\cal R}({\bf
Q}\,).
\end{eqnarray}
Then, using the one-dimension imaging technique
\cite{DAB97,DAB98,DAB01} one can get
$S^l_{\alpha_1,\cdots,\alpha_l}(r)$ by Eq. (\ref{RS}), and finally
get the three-dimension source function $S({\bf r})$ with Eq.
(\ref{S3r}).

\begin{figure} [h]
\vspace*{6mm}
\includegraphics[scale=0.5]{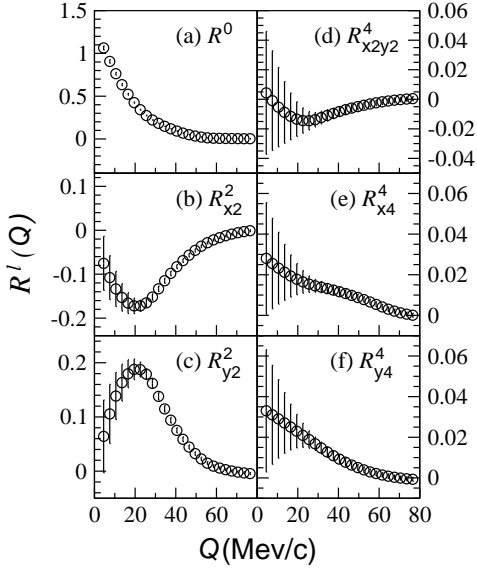}
\vspace*{8mm} \caption{The independent
$R^l_{\alpha_{\!1},\cdots,\,\alpha_l}(Q)$ up to $l=4$ for the
granular source.} \label{FigRxyz}
\end{figure}

In Fig. \ref{FigRxyz} we show the independent
$R^l_{\alpha_{\!1},\cdots,\alpha_l}(Q)$ up to $l=4$ for the granular
source.  Here $R^0(Q)$ is the correlation function averaged over
angles, the index ``$\alpha_i2$", ``${\alpha_i4}$", and
``${\alpha_i2\alpha_j2}"$ denote ``${\alpha_i,\alpha_i}$",
``${\alpha_i,\alpha_i,\alpha_i,\alpha_i}$", and
``${\alpha_i,\alpha_i,\alpha_j,\alpha_j}$", respectively.  Because
of source symmetry, the components with odd $l$ are zero.  We use
$x$ and $y$ denote the directions parallel and perpendicular to the
transverse momentum of the pion pair, and let $z$-axis along the
longitudinal (beam) direction.  The $x$, $y$, and $z$ directions are
usually called as ``out", ``side", and ``long" directions in HBT
interferometry \cite{GBE88,SPR90}.

The source function along the $j$-axis ($r=r\!_j$, $j=x,\,y,\,{\rm
or}\,z$) can be expressed as \cite{PDA05,PDA07,PHE08}
\begin{equation}
S(r\!_j)=S^0(r\!_j)+S^2_{j\,2}(r\!_j)+S^4_{j\,4}(r\!_j)+\cdots,
\end{equation}
where $S^0$ is the angle-averaged source function and
$S^l_{j\,l}~(l=2,4,\cdots)$ are the $l$-order modifications for
$S^0$ in the $j$-direction.

\begin{figure} [h]
\includegraphics[scale=0.65]{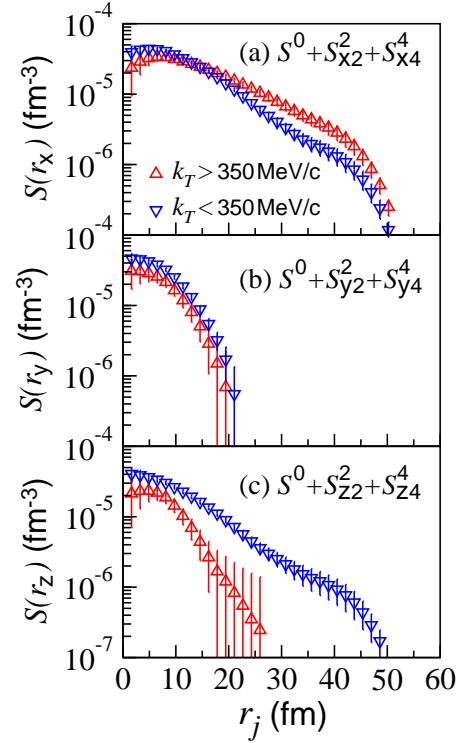}
\vspace*{2mm} \caption{(Color online) The three-dimension source
functions of the granular source for $k_T < 350$ MeV/c and $k_T >
350$ MeV/c. $r\!_j=r_x,\,r_y,\,\mbox{or}~r_z$. } \label{FigS12}
\end{figure}

\begin{figure} [h]
\includegraphics[scale=0.46]{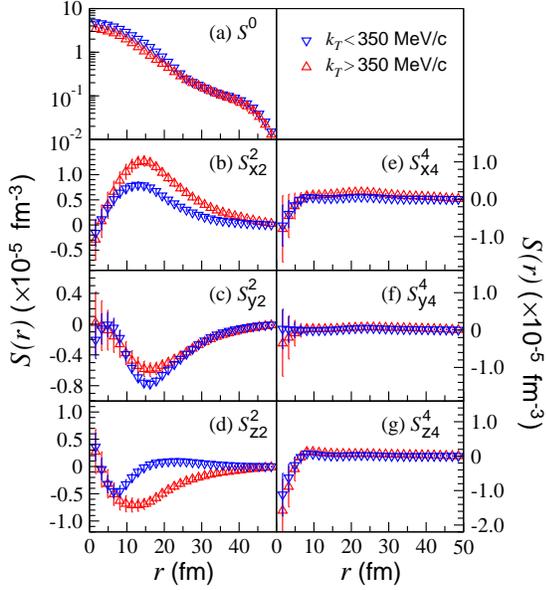}
\vspace*{2mm} \caption{(Color online) The zero-order, second-orde,
and fourth-order source functions $S^0(r)$, $S^2(r)$, and $S^4(r)$
in ``out", ``side", and ``long" directions for $k_T < 350$ MeV/c and
$k_T > 350$ MeV/c for the granular source. } \label{FigSxyz}
\end{figure}

In Fig. \ref{FigS12} we show the three-dimension source functions of
the granular source for $k_T<350$ MeV/c and $k_T>350$ MeV/c, where
$k_T=|{\bf p}_{1T}+{\bf p}_{2T}|/2$ is the transverse momentum of
pion pair in the longitudinally comoving system (LCMS) \cite{MAL05}.
One can see that the widths of the source functions in ``out"
direction are larger than those in ``side" direction as observed in
experiments \cite{PHE08}.  This reflects that the source expansion,
which boosts the pair velocity (momentum), leads to different
geometries in ``out" and ``side" directions.  For higher $k_T$, the
width of the source function in ``out" direction is larger than that
for lower $k_T$.  However, in ``side" direction the width of the
source function are smaller for higher $k_T$ than that for lower
$k_T$. The source function in longitudinal direction has a long tail
for the pion pairs with lower $k_T$ for the granular source.  It is
because that the average longitudinal velocity of pion pairs is
larger for smaller $k_T$.  Our model calculations indicate that the
average velocity of the pair in longitudinal direction is about 0.43
for $k_T<350$ MeV/c and 0.35 for $k_T>350$ MeV/c.  In Fig.
\ref{FigSxyz} we show the angle-averaged source functions $S^0(r)$
and the ``second-order" and ``fourth-order" source functions
$S^2(r)$ and $S^4(r)$ in ``out", ``side", and ``long" directions for
$k_T < 350$ MeV/c and $k_T > 350$ MeV/c for the granular source.  It
can be seen that the one-dimension imaging $S^0$ of the granular
source exhibits a ``two-tiered" structure \cite{ZTY09}.  The
fourth-order source functions are almost zero exception for that at
small $r$.

In order to examine the three-dimension source functions
quantitatively, we introduce the moments of $r_i$ for the
``$i$-component" of the source function,
\begin{eqnarray}
\label{rin}
\langle r_i^n\rangle=\frac{\int_0^\infty
d\,r_i\,r_i^n\,S(r_i)}{\int_0^\infty d\,r_i\,S(r_i)}\,,
~~~n=1,2,\cdots,
\end{eqnarray}
and define the normalized first-order moment and deviation as
\begin{equation}
\label{wRi} \widetilde{R}_i \equiv \frac{\sqrt{\pi}}{2}\langle
r_i\rangle
\end{equation}
and
\begin{equation}
\label{wsigmai} \widetilde{\sigma}_i \equiv \sigma_i
/\sqrt{2-4/\pi}\,,
\end{equation}
which are normalized to the Gaussian radius $R_{\rm ga}$ for a
one-dimension Gaussian source [$S(r_i)\sim \exp(r^2_i/4R_{\rm ga})$].

\begin{figure} [h]
\includegraphics[scale=0.55]{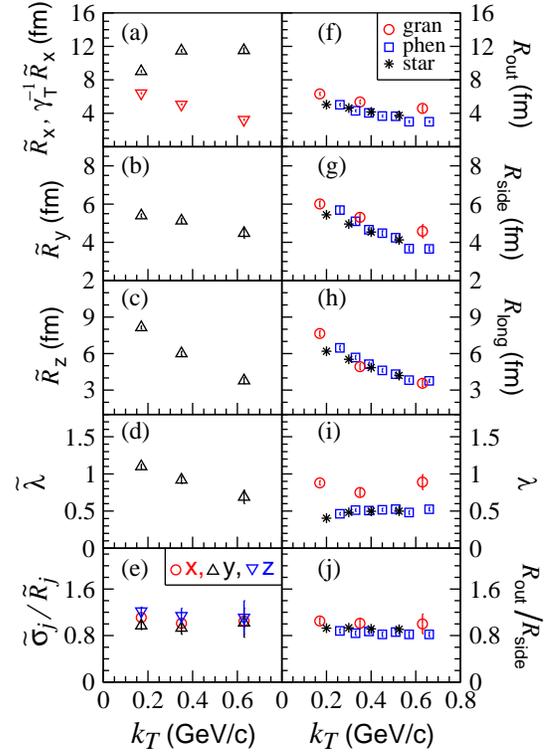}
\vspace*{7mm} \caption{(Color online) (a)--(e) The characteristic
quantities of three-dimension imaging as functions of $k_T$ for the
granular source.  (f)--(j) The HBT Gaussian fitted results for the
granular source compared with the results of RHIC experiments
\cite{PHE04a,STA05a}.} \label{FigHBTr3}
\end{figure}

In Fig. \ref{FigHBTr3} (a), (b), and (c) we exhibit the normalized
first-order moments (symbols \!{\raisebox{-1.2mm}{\textsuperscript{
$\triangle$}}}) of the granular source as functions of $k_T$ in
``out", ``side", and ``long" directions, respectively.
$\widetilde{R}_x$ increases and $\widetilde{R}_y$ decreases with
$k_T$ as expected.  Because the average longitudinal velocity of the
pairs with larger $k_T$ is smaller, $\widetilde{R}_z$ decreases with
$k_T$.  Figure \ref{FigHBTr3} (d) gives the results of
$\widetilde{\lambda}$ calculated with the angle-averaged source
function $S^0(r)$ in Eq. (\ref{wl}) for the granular source.  The
results of $\widetilde{\lambda}$ decrease with $k_T$.  In Fig.
\ref{FigHBTr3} (e) we show the ratios $\widetilde{\sigma}_j/
\widetilde{R}_j~(j=x,\,y,\,z)$ for the granular source.  The large
value of $\widetilde{\sigma}_z/\widetilde{R}_z$ at small $k_T$
indicates that the source function have a serious deviation from
that of a Gaussian source in longitudinal direction.  However, in
``side" direction the ratios are almost unit.

The first-order moments $\widetilde{R}_x$, $\widetilde{R}_y$, and
$\widetilde{R}_z$ describe the average separations of the source in
PCMS.  In LCMS the pair has a transverse velocity $v_T$
($=k_T/[(E_1+E_2)/2]$).  The spatial separation in LCMS in ``out"
direction is smaller than $\widetilde{R}_x$ by the Lorentz
contraction factor $\gamma^{-1}_T=\sqrt{1-v^2_T}$ \cite{SPR90}.  In
Fig. \ref{FigHBTr3} (a), the symbols
{\raisebox{-1.2mm}{\textsuperscript{ $\nabla$}}} denote the results
of $\gamma^{-1}_T\widetilde{R}_x$.  One can see that $\gamma^{-1}_T
\widetilde{R}_x$ decrease with $k_T$ as that of the HBT radius
$R_{\rm out}$ in LCMS \cite{CYW94,UAW99,RMW00,MAL05}.  For
comparison, in Fig. \ref{FigHBTr3} (f) -- (j) we exhibit the HBT
radii and chaotic parameter for the granular source fitted with
\begin{equation}
C(q_{\rm out}, q_{\rm side}, q_{\rm long})=1+\lambda \,e^{-q^2_{\rm
out} R^2_{\rm out} -q^2_{\rm side} R^2_{\rm side} -q^2_{\rm long}
R^2_{\rm long}},
\end{equation}
in LCMS and the experimental HBT results of PHENIX \cite{PHE04a} and
STAR \cite{STA05a} for $\sqrt{s_{\rm NN}}=200$ GeV Au+Au collisions.
One can see that the HBT radii as functions of $k_T$ for the
granular source agree with the experimental results.  Also, the
separations $\gamma^{-1}_T \widetilde{R}_x$, $\widetilde{R}_y$, and
$\widetilde{R}_z$ as functions of $k_T$ agree with those of the
corresponding HBT radii $R_{\rm out}$, $R_{\rm side}$, and $R_{\rm
long}$.  At lower $k_T$ the values of the chaotic parameter
$\lambda$ are smaller than those of $\widetilde{\lambda}$ obtained
by imaging analysis, and both the values of $\lambda$ and
$\widetilde{\lambda}$ of the granular source are larger than the
experimental $\lambda$ results.  The results of $R_{\rm out}/R_{\rm
side}$ for the granular source are consistent with the experimental
data.  The granular source model reproduce the main characteristics
of the source functions as well as HBT radii of RHIC experimental
measurements \cite{PHE04a,STA05a,PHE08}.

\section{Summary and Conclusion}

Imaging analysis is a model-independent technique
\cite{DAB97,DAB98,DAB01}.  With the source functions obtained by the
three-dimension imaging technique \cite{PDA05,PDA07}, one can
calculate numerically the first-, second-, and even higher-order
moments of the spatial separation $r$ in different directions. In
principle, the detailed information about the source geometry,
coherence, and dynamics can be extracted by analyzing these moments.

In this paper we examine the spatial and coherent information of
pion-emitting source extracted by usual HBT Gaussian fit and imaging
analysis in relativistic heavy ion collisions.  The usual HBT
results of the Gaussian fit are model-dependent.  They are
inappropriate for describing the characteristics of the sources with
non-Gaussian distributions, such as the sources with granular and
core-halo structures.  However, the zero- and first-order moments as
well as the deviation of two-pion source separation
($\widetilde{\lambda}, \widetilde{R}, \widetilde{\sigma}$) obtained
by imaging analysis can provide better descriptions for the source
coherent and spatial characteristics.  They are model-independent
characteristic quantities of the particle-emitting sources.

Based on an improved granular source model we investigate the
characteristic quantities of the pion-emitting source produced in
relativistic heavy ion collisions.  We find that the granular source
model of QGP droplets can reproduce the main characteristics of the
experimental two-pion correlation functions and source functions in
$\sqrt{s_{\rm NN}}=200$ GeV Au+Au collisions
\cite{PHE04a,STA05a,PHE08}.  In the transverse directions of the
collisions, the width of the source function in ``out" direction is
larger than that in ``side" direction.  Correspondingly, the value
of the first-order moment $\widetilde{R}_x$ is larger than that of
$\widetilde{R}_y$.  The dependence of $\widetilde{R}_j$ on the
transverse momentum of the pair $k_T$ exhibits different in ``out"
and ``side" directions. $\widetilde{R}_x$ increases but
$\widetilde{R}_y$ decreases with $k_T$ increase.  In the
longitudinal direction, the source function for small $k_T$ has a
long tail, which is much different from the Gaussian source
function.  Correspondingly, the value of
$\widetilde{\sigma}_z/\widetilde{R}_z$ is large at small $k_T$.  We
find that $\widetilde{R}_z$ decrease rapidly with $k_T$ increase.
After taking into account the Lorentz contraction in ``out"
direction, all the quantities $\gamma^{-1}_T \widetilde{R}_x$,
$\widetilde{R}_y$, and $\widetilde{R}_z$ decrease with $k_T$
increase.  These transverse-momentum dependences for the granular
source are consistent with those of the usual HBT Gaussian fit
results of $R_{\rm out}$, $R_{\rm side}$, and $R_{\rm long}$.  They
are also consistent with the results of RHIC experiments
\cite{PHE04a,STA05a}.

From our model results we notice that both the values of
$\widetilde{\lambda}$ and $\lambda$, obtained by the imaging
analysis and HBT Gaussian fit, are larger than the experimental
results of $\lambda$ \cite{PHE04a,STA05a}.  Except for the source
coherence, there are other elements which may affect the results of
$\widetilde{\lambda}$ and $\lambda$, such as the Coulomb interaction
of the final-state particles and source dynamics
\cite{SPR86,CYW94,UAW99,RMW00,MAL05}.  Investigating the effects of
these elements on $\widetilde{\lambda}$ and $\lambda$ for the
granular source will be an interesting issue.  On the other hand,
because imaging analysis is performed in PCMS the temporal
information of the particle emission is hidden \cite{DAB01,SPR86}.
Further investigation on how to extract the temporal information of
the source by analyzing the spatial moments in different directions
and their $k_T$ dependences will be also of interest.

\begin{acknowledgments}
This research was supported by the National Natural Science
Foundation of China under Contract No. 10775024.
\end{acknowledgments}

\end{document}